\documentclass[conference]{IEEEtran}
\usepackage{times}
\usepackage{soul}
\usepackage{url}
\usepackage[hidelinks]{hyperref}
\usepackage[utf8]{inputenc}
\usepackage[small]{caption}
\usepackage{graphicx}
\usepackage{amsmath}
\usepackage{amsthm}
\usepackage{booktabs}
\usepackage{algorithm}
\usepackage{algorithmic}
\usepackage[switch]{lineno}
\usepackage[table,xcdraw]{xcolor}
\usepackage{multirow}

\usepackage{float}

\ifCLASSINFOpdf
\else
\fi
\begin{document}
%
\title{A Survey on Model-heterogeneous Federated Learning: Problems, Methods, and Prospects}




%

\author{\IEEEauthorblockN{Boyu Fan\IEEEauthorrefmark{1},
Siyang Jiang\IEEEauthorrefmark{2},
Xiang Su\IEEEauthorrefmark{3},
Sasu Tarkoma\IEEEauthorrefmark{1} and
Pan Hui\IEEEauthorrefmark{4}\IEEEauthorrefmark{1}}
\IEEEauthorblockA{\IEEEauthorrefmark{1}University of Helsinki, Finland}
\IEEEauthorblockA{\IEEEauthorrefmark{2}The Chinese University of Hong Kong, Hong Kong}
\IEEEauthorblockA{\IEEEauthorrefmark{3}Norwegian University of Science and Technology, Norway}
\IEEEauthorblockA{\IEEEauthorrefmark{4}The Hong Kong University of Science and Technology (Guangzhou), China}
}


\maketitle

\begin{abstract}
As privacy concerns continue to grow, federated learning (FL) has gained significant attention as a promising privacy-preserving technology, leading to considerable advancements in recent years. Unlike traditional machine learning, which requires central data collection, FL keeps data localized on user devices. However, conventional FL assumes that all clients operate with identical model structures initialized by the server. In real-world applications, system heterogeneity is common, with clients possessing varying computational capabilities. This disparity can hinder training for resource-limited clients and result in inefficient resource use for those with greater processing power. To address this challenge, model-heterogeneous FL has been introduced, enabling clients to train models of varying complexity based on their hardware resources. This paper reviews state-of-the-art approaches in model-heterogeneous FL, analyzing their strengths and weaknesses, while identifying open challenges and future research directions. To the best of our knowledge, this is the first survey to specifically focus on model-heterogeneous FL.
\end{abstract}

%
\IEEEpeerreviewmaketitle


\section{Introduction}
In the information era, vast amounts of data are continuously generated at an unprecedented scale, driven by sensors and human activities. Combined with machine learning techniques, this big data brings significant benefits to society, influencing our lives in numerous ways. However, growing concerns about privacy issues arise from the need to collect data from various parties for model training. With the introduction of various privacy protection policies, federated learning (FL) has emerged as a promising solution for conducting model training in a privacy-preserving manner~\cite{mcmahan2017communication}. FL eliminates the need for centralized data aggregation by allowing data to remain stored locally on clients’ devices, ensuring that private information never leaves its original environment. Each client individually trains the model with its local data and contributes to the global model by only transmitting model parameters or gradients to the central server. Due to this privacy-preserving nature, FL has been widely adopted in diverse domains such as healthcare~\cite{rauniyar2023federated}, finance~\cite{10342702}, and transportation~\cite{kumar2023federated}. 

Conventional FL methods rely on the assumption that all participating clients have homogeneous system resources~\cite{mcmahan2017communication,10.1145/3679013,9252927}. In other words, they presume that all clients possess the hardware capabilities necessary for completing local training. However, in the real world, clients often often have varied computational capabilities, leading to system heterogeneity~\cite{Liu_Jia_2024}. Moreover, training a neural network is computationally expensive, given a huge number of model parameters and the necessity for multiple iteration rounds. When clients with limited computational power fail to complete local training within the expected timeframe, it creates straggler issues, forcing both the server and other clients to wait, ultimately leading to inefficient resource utilization~\cite{10364357,10529949}. 

A natural solution is to enable clients to train models of varying complexity according to their resources. To this end, model-heterogeneous FL~\cite{10633723,fan2024fedtsaclusterbasedtwostageaggregation} has been investigated to solve the aforementioned challenge, alleviating the requirement imposed by traditional FL where all clients are expected to share a universal global model. Instead, it enables clients to train different types of models based on their respective capabilities, which offers two significant advantages, i.e., mitigating the straggler issues and improving resource utilization efficiency. For instance, envision a situation where a Raspberry Pi can efficiently train a two-layer convolutional neural network (CNN), while a ResNet18 model~\cite{he2016deep} would pose a significant burden on it. Lengthy training times result in waiting periods for other clients, leading to idle periods for other clients and inefficient resource utilization. In contrast, model-heterogeneous FL allows a Raspberry Pi to train a small model, while a server machine can simultaneously train a ResNet18 model within the same FL round. 

Despite the promise of model-heterogeneous FL, a significant challenge remains: how to aggregate models with different architectures? Directly averaging model parameters is infeasible due to structural differences among models. Therefore, innovative approaches are required to enable effective aggregation under these heterogeneous conditions. This survey offers a comprehensive review of state-of-the-art model-heterogeneous FL techniques, categorizing them into two main approaches: knowledge distillation (KD)-based methods, partial training (PT)-based methods, along with several techniques that do not fall into these categories. We analyze the strengths and limitations of each approach, discuss current challenges, and propose future research directions. To the best of our knowledge, this is the first survey dedicated to model-heterogeneous FL, representing a significant contribution to this emerging field. In contrast, ~\cite{ye2023heterogeneous} touches only briefly on model heterogeneity, focusing more broadly on generalized heterogeneity in FL. Given that numerous existing works have already addressed the challenges of data heterogeneity, this survey does not focus on methods for solving data heterogeneity. Instead, it concentrates on the issue of model heterogeneity arising from system heterogeneity.

The main contributions are summarized as follows:
\begin{itemize}
    \item We present the first survey dedicated to model-heterogeneous FL, summarizing state-of-the-art methods and categorizing them into two main strategies: knowledge distillation (KD)-based and partial training (PT)-based methods, along with several novel approaches that do not fit into these categories.
    \item We analyze the advantages and limitations of each method, providing readers with a clear comparison to facilitate understanding and application of these works.
    \item We highlight open problems and propose future research directions in model-heterogeneous FL, enabling researchers to quickly identify key challenges and offering insights for further exploration.
\end{itemize}

\section{Model-heterogeneous FL: Fundamental concepts and challenges}
\label{background}
In this section, we introduce the fundamental concepts related to FL and discuss the challenges posed by heterogeneity. We then provide an overview of model-heterogeneous FL to lay the groundwork for the subsequent sections.

\subsection{Federated Learning}
\begin{figure}[h]
    \setlength{\abovecaptionskip}{-0.cm}
    \setlength{\belowcaptionskip}{-0.cm}
    \centering
    \includegraphics[width=0.4\textwidth]{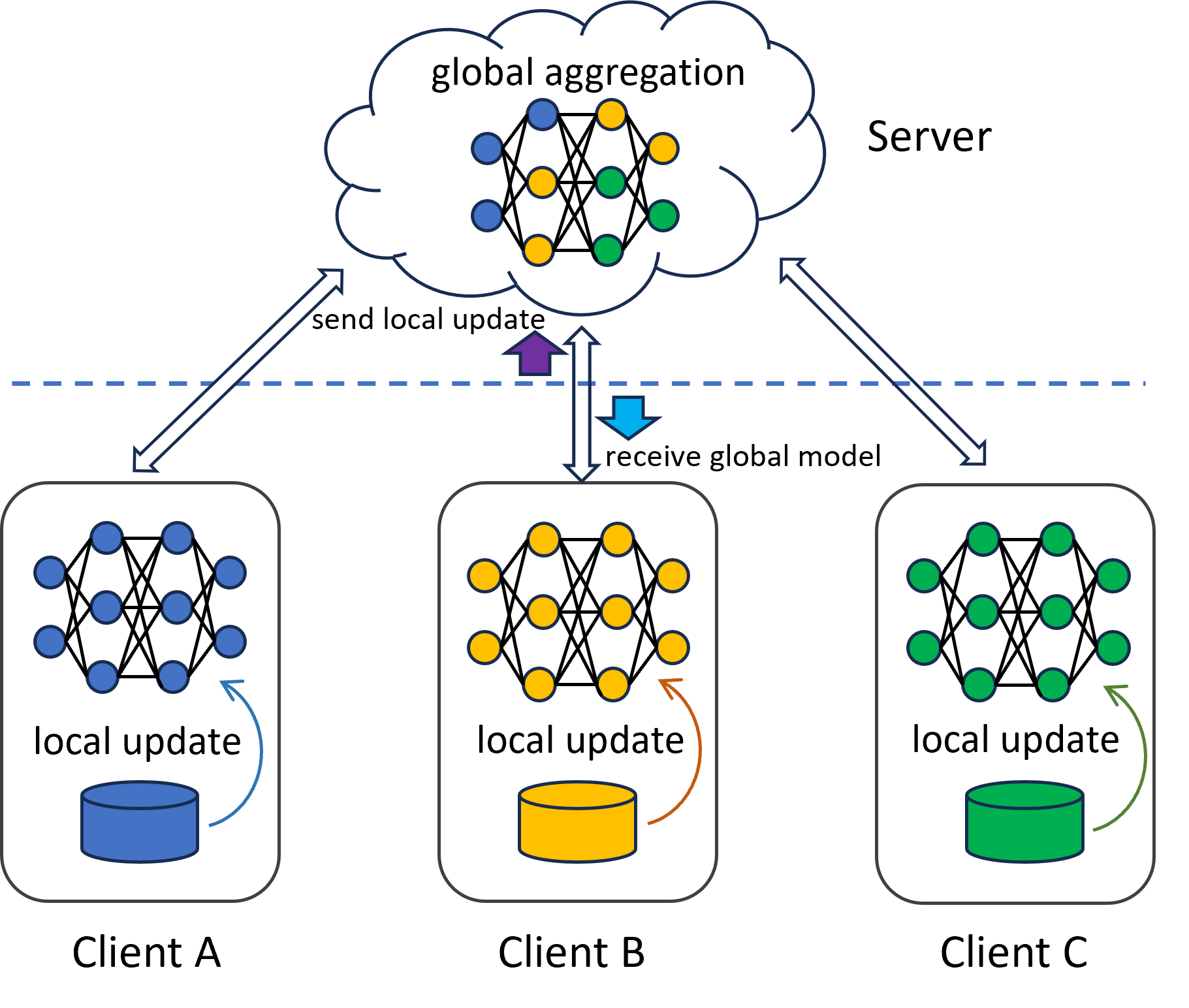}
    \caption{A standard FL architecture.}
    \label{fl_generalarchitecture}   
\end{figure}

FL is first proposed by Google~\cite{mcmahan2017communication} and becomes a paradigm to train ML models without the central collection of data. Since then, ongoing research has developed lots of new applications for FL. Figure~\ref{fl_generalarchitecture} presents the general FL architecture, consisting of two main entities, i.e., a central server for conducting aggregation and multiple clients for performing local training. Each client trains models with its local dataset without sharing with others, thereby ensuring the protection of their privacy. During the training process, instead of collecting raw data from different sources, the server only receives model parameters from clients and sends the aggregated model parameters back to the clients to update the local models. 

In general, the FL training process includes three main steps: 1) Server initialization. As the first step of FL training, the server initializes a global model tailored to the specific training task and subsequently distributes the model to the selected clients. 2) Local Update. Each selected client $k$ conducts model training with its local data $D_{k}$ based on the received global model. Once the training is completed, the clients send the updated models back to the server. 3) Model aggregation. The server receives the updated models from clients and then conducts model averaging aggregation to update the model. The aggregated model will become the new global model and will be iteratively updated in subsequent rounds until convergence.


\subsection{System Heterogeneity}
System heterogeneity refers to the diversity in hardware configurations, software platforms, communication protocols, and even run-time environments among different participants and devices. The traditional FL workflow typically involves the server waiting to receive updated parameters from the selected clients before performing aggregation. However, due to the existence of system heterogeneity, different clients may present varying training efficiency, leading to two potential problems. 
First, due to the fact that all clients perform local updates based on a unified global model, the required training times vary based on their computation capabilities. Consequently, both the server and the clients with more abundant computation resources need to wait for slower clients, resulting in resource wastage. 
Second, some resource-constrained devices are not capable of training complex neural networks, e.g., ResNet. Extremely long processing time can cause the server to mistakenly assume that the clients are offline, thus never selecting them for the distributed training. As the local data owned by these devices cannot join the training, the contribution to the global model can be disproportionately small, leading to imbalanced models. 
In a nutshell, system heterogeneity poses challenges to the performance, generalization, coordination and efficiency of the FL system. 

Notably, most existing work on FL focuses on addressing data heterogeneity, while papers addressing system heterogeneity constitute a relatively small proportion. However, both of them are essential to consider when FL is deployed in real-world applications. 

\subsection{Model-heterogeneous Federated Learning}
The optimal approach for addressing system heterogeneity lies in empowering devices to learn diverse models with varying complexities based on their available resources. 
For example, training a CNN with two convolutional layers requires significantly fewer computing resources compared to training a ResNet18. Therefore, model-heterogeneous FL, allowing clients to train heterogeneous models, naturally emerges as a promising technique for mitigating system heterogeneity issues. Specifically, each client can choose a model that suits its own capabilities, rather than being required to use a uniform global model initialized by the server. This approach can significantly enhance the efficiency of local training, effectively mitigating issues such as stragglers and resource wastage.

\begin{figure}[h]
    \setlength{\abovecaptionskip}{-0.cm}
    \setlength{\belowcaptionskip}{-0.cm}
    \centering
    \includegraphics[width=0.45\textwidth]{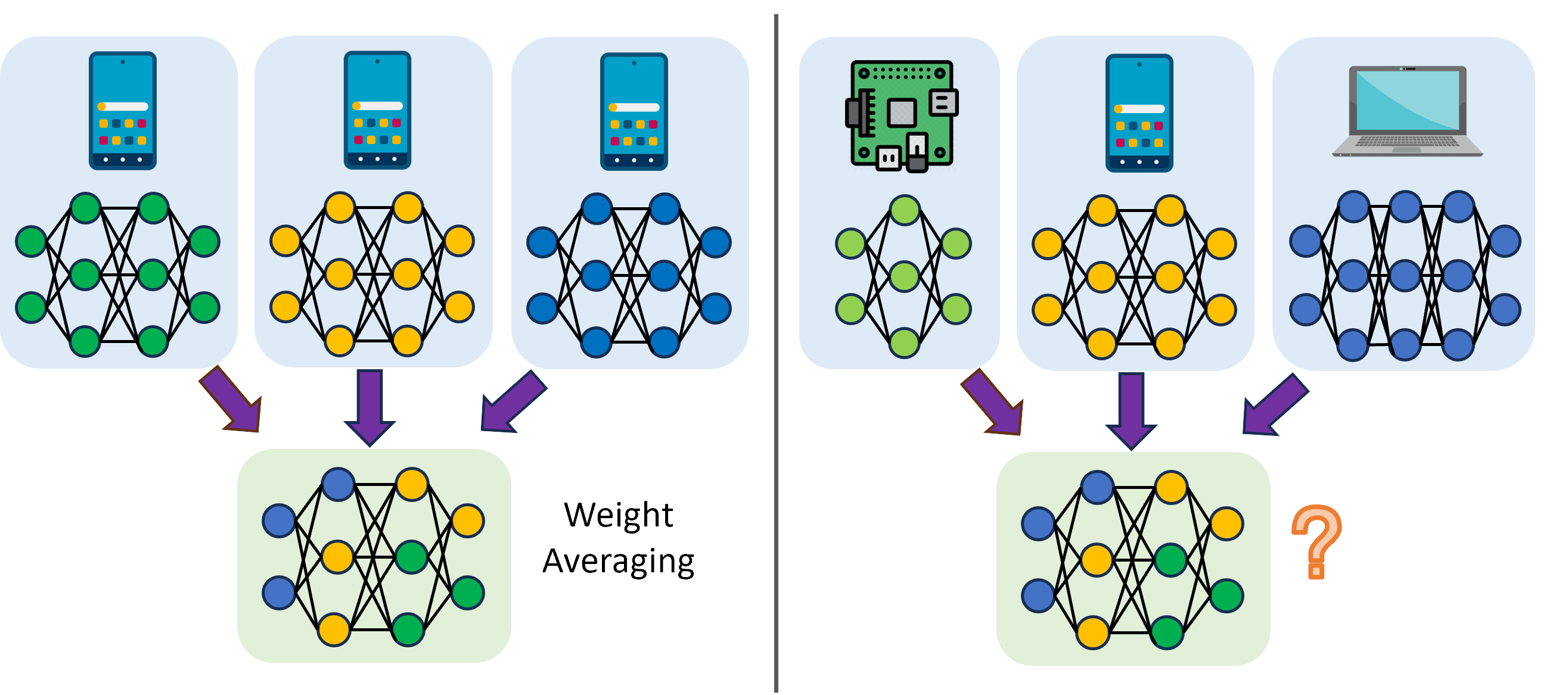}
    \caption{The overview of the challenge in model heterogeneity.}
    \label{hetero_aggre}   
\end{figure}  

However, in model-heterogeneous FL, clients may possess different model architectures, thereby complicating the aggregation process on the server side. Figure~\ref{hetero_aggre} illustrates this problem. In traditional FL, there exists only one global model architecture that is shared by all clients, facilitating straightforward element-wise averaging operations on multiple model parameters. In contrast, the heterogeneous models have varying structures, leading to different shapes of weight parameters thus preventing vanilla aggregation. Therefore, how to achieve effective model aggregation for models with different architectures becomes the core challenge of model-heterogeneous FL. 

\section{State-of-the-art Model-heterogeneous FL Approaches}
\label{model-heterogeneous}
In this section, we investigate the existing approaches for model-heterogeneous FL, which can be classified into two main solutions, i.e., KD-based approaches and PT-based approaches. We also present several alternative approaches that fall outside of these two categories. A summary of them is listed in Table~\ref{summary} at the end.

\subsection{KD-based Approaches}

\subsubsection{Knowledge Distillation}
KD was proposed by~\cite{hinton2015distilling}, which aims to distill knowledge from a larger model or an ensemble of models and transfer this understanding to a smaller, more compact model. Despite its reduced size, this distilled model, often referred to as the ``student" model, can exhibit performance metrics remarkably similar to its larger ``teacher" counterpart. 
During the KD process, three types of knowledge play pivotal roles, including response-based, feature-based, and relationship-based knowledge~\cite{gou2021knowledge}. In response-based knowledge, information is extracted from the output layer of the teacher model, capturing the class probabilities. Feature-based knowledge, on the other hand, does not restrict itself to the output of the last layer, it also gathers insights from the outputs of intermediate layers of a neural network. Contrasting with the prior two types, relationship-based knowledge leverages the relationships either between different layers of a neural network or between individual data samples. Various techniques, such as similarity matrices, instance relationship graphs, or mutual information flows, have been proposed to define these relationships.

\subsubsection{Knowledge Distillation in FL}
The output of the last layer of a model, known as logits in response-based knowledge, has a shape determined by the number of classes in the dataset, which is independent of the model's architecture. For instance, in the CIFAR-10 dataset, the logits' shape remains at $10 \times 1$, whereas in the CIFAR-100 dataset, it is $100 \times 1$. This characteristic allows the utilization of KD in FL, enabling aggregation at the server even when received models have varying architectures. 

\begin{figure}[h]
    \setlength{\abovecaptionskip}{-0.cm}
    \setlength{\belowcaptionskip}{-0.cm}
    \centering
    \includegraphics[width=0.4\textwidth]{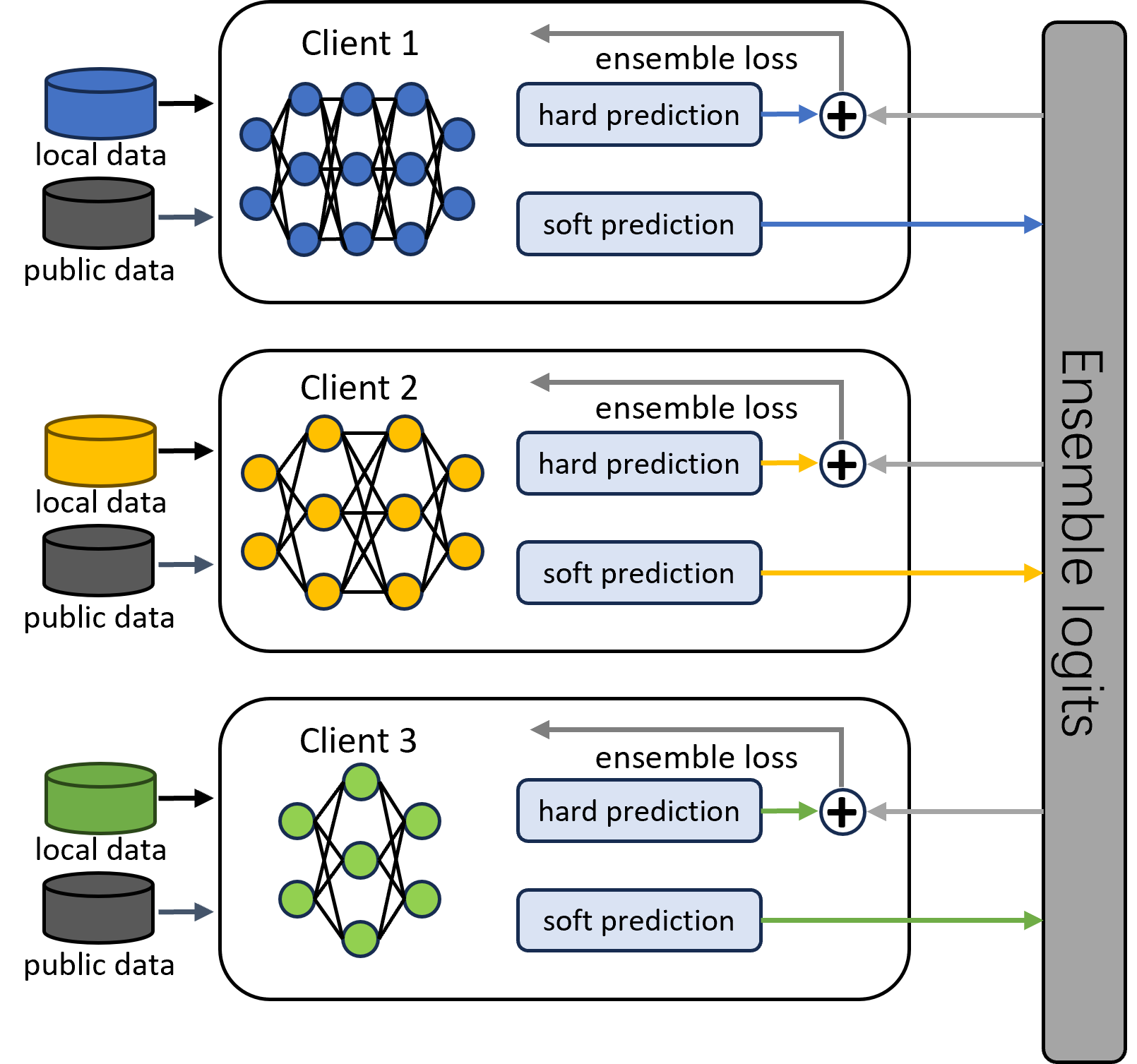}
    \caption{The illustration of KD in the model-heterogeneous FL.}
    \label{KD-based-FL}   
\end{figure}  

In recent years, several approaches have been developed to apply KD to achieve model-heterogeneous FL. The basic idea is presented in Figure~\ref{KD-based-FL}. Clients use their local data to conduct hard predictions and leverage public data to serve as soft predictions. FedMD~\cite{li2019fedmd} is a pioneering work to achieve model aggregation using KD rather than averaging the model parameters. The training process of FedMD consists of four steps: First, each client computes the class scores on a public dataset and transfers them to the central server. Second, the server computes the average logits to serve as a consensus and sends it to all clients. Third, each client trains its model to approach the consensus on the public dataset. Finally, each client trains its model based on the local data to achieve local updates. Compared to conventional FL approaches, FedMD does not need to exchange model parameters between server and client, allowing models to have different architectures. Clients contribute to the global model by only sharing their logits that are computed on the public dataset. 

Although FedMD bypasses model parameter aggregation, its global logit construction by simple averaging at the server may not optimally harness the collective knowledge of clients. KT-pFL~\cite{zhang2021parameterized} introduces a knowledge coefficient matrix in the FL training process. After training the clients' personalized models over the local data, the local model logits are generated to the public dataset and shared with the server. Then, using the linear combination of the aggregated soft predictions and the knowledge coefficient matrix, the server calculates the local soft prediction for each client. After each iteration, the knowledge coefficient matrix is updated on the server. With the help of a knowledge coefficient matrix, each client can adaptively aggregate all local soft predictions to form a personalized one instead of being restricted to a global soft prediction.

A shared limitation of the aforementioned approaches is their dependency on a public dataset. In practical applications, this kind of public dataset requires careful deliberation and even prior knowledge of clients' private data. To eliminate this dependency, several approaches for data-free KD have been proposed. Compared to FedMD, FedDF~\cite{lin2020ensemble} relieves the requirement of labeled public data. Instead, unlabeled or artificially generated data from arbitrary domains is used to perform ensemble distillation on the server. This involves training local models with unlabeled or generated data and using the resulting output logits to construct the global model. In contrast to FedMD, distillation is carried out on the server side, leaving the local training unaffected. 

Similarly, FedGen~\cite{pmlr-v139-zhu21b} achieves data-free learning by training a generative model based on the prediction rules of the local clients. Through this approach, the generator effectively models the global data distribution. With this distilled joint knowledge about data distribution from the generator, local models are then updated. This methodology allows the local models to mutually benefit from each other without the necessity of employing an external public data source. A notable distinction from earlier frameworks is that the distilled knowledge is not passed on to a global model, but directly to the local models. However, a potential drawback of FedGen pertains to its tendency to lose information from the local models, due to the fact that the approach sometimes overlooks the incompatibility of localized knowledge during the distillation process.

FedZKT~\cite{9912292} presents an alternative approach to data-free KD for FL. In contrast to FedGen, FedZKT keeps the computation-heavy generator on the server side to relieve devices with low computational capacity. The on-device parameters are not shared with the server to preserve privacy. The generative model provides inputs for global and on-device training, which are utilized to detect and maximize the disagreements between the global and private on-device models. Moreover, the pre-trained generative model is also utilized to distill the on-device model and update its parameters with global knowledge. The approach of alleviating clients from some computation-heavy tasks results in an increase in the number of classes per client in comparison to FedMD. 

FedFTG~\cite{Zhang_2022_CVPR} offers a solution by fine-tuning the global model through a process of hard sample mining. This method harnesses the robust computational capacities of the server effectively. FedFTG employs a generator to extract knowledge from the input spaces of individual local models. Central to this process is the utilization of a hard sample mining scheme, prompting the generator to produce challenging samples within the data distribution that highlight disparities between the local and global models. This approach ensures the global model is finely attuned to the overall data distribution. In addition, the framework integrates customized label sampling and a class-level ensemble strategy to offset variances in label distributions among local clients resulting from class imbalances and varying importance levels of knowledge for the classes. 

Another generator-based work is FedTSA~\cite{fan2024fedtsaclusterbasedtwostageaggregation}. Different from the above methods, FedTSA innovatively adapts a diffusion model on the server side to improve the data generation quality, thereby achieving higher performance. Specifically, this method first conducts clustering to group clients with varying processing abilities, then performs deep mutual learning based on the data generated by the diffusion model to achieve knowledge sharing among heterogeneous models. 

Notably, in some cases, the data generator may not perform well if the client data is highly non-IID, making the generated data deviate from the real distribution. With the increase in FL rounds, the output distribution may undergo large shifts, which highly decreases the KD process. To this end, DFRD~\cite{wang2024dfrd} trains a conditional generator that consists of three loss: fidelity losses, transferability and diversity, to make sure the generated data can match the real data. Further, this method equips the server with an exponential moving average copy of the generator, aiming to mitigate catastrophic forgetting during the FL process. 

In FedHKD~\cite{chen2023best}, the hyper knowledge is introduced as the average of the data representation of a class in combination with the respective average soft prediction. The iterative process unfolds as follows: 1) Each client sends the local hyper-knowledge of each class to the server. To safeguard the client's privacy, noise is intentionally added to the data representation. 2) The global hyper knowledge is aggregated from the collected hyper knowledge of the clients. The added noise is compensated during the aggregation. 3) The global hyper knowledge is sent back to the clients, who adjust their loss function accordingly for the subsequent iteration. In comparison to FedMD, FedHDK is competitive in local and global test accuracy for heterogeneous data. 

Building on the success of using response-based KD on FL, researchers believe the logits implicitly possess the label information of each class, serving as a channel for knowledge transfer among different parties. FedProto~\cite{tan2022fedproto} capitalizes on this idea by employing class prototypes as a medium for knowledge exchange. Specifically, a class prototype is a representation of a class, derived from the embedding vectors of class features and subsequently averaged. These prototypes are sent to the server, where they are aggregated to create a global knowledge base. The global knowledge is then sent back to the clients, motivating them to refine their local prototypes, which in turn bolsters the efficacy of local training. In this way, no global model is maintained, allowing clients to employ different model architectures. The following work DeProFL~\cite{10246848} applies the idea of the prototype on Internet of Things to decrease the communication cost and enable model-heterogeneous FL training.


FedGH~\cite{10.1145/3581783.3611781} enables model-heterogeneous FL by separating the model into a heterogeneous feature extractor and a homogeneous prediction head. Specifically, each client trains a local model and sends the representations generated by its feature extractor, along with class labels, to the server. The server then uses this data to train a global prediction head, which is subsequently sent back to the clients to replace their local heads. The global head aggregates knowledge from all clients, acting as a bridge for knowledge transfer. Unlike other approaches, FedGH does not require a public dataset for representation alignment since the uploaded data is already labeled. However, this approach has a notable drawback: the generated representations are closely tied to the original data, posing a potential privacy risk.

~\cite{regatti2022conditional} investigates a solution to preserve the model privacy of clients in addition to data privacy. The Fed-CMA algorithm allows clients to maintain individual model architectures while participating in FL with other clients. This approach does not require a public dataset or maintain a global model. Instead, the feature weights of the feature extractor and the classification weights are distilled client-side. The server then aggregates the weights from all clients and sends the aggregated feature and classification weights back to the clients to update the local model. To achieve this, a new algorithm based on conditional moment alignment and generalization for aggregating the weights is introduced, since conventional algorithms such as FedAvg and FedProx cannot be used due to the different client model architectures.

\subsubsection{Limitations}
The KD-based concept facilitates various kinds of new FL approaches to solve the model-heterogeneous FL problem. However, several limitations exist. First, despite several claims of achieving data-free KD, they still require some kind of data that the server and client can jointly access to achieve knowledge alignment. How to achieve truly data-free KD remains an interesting yet challenging question. Second, most of the existing works focus on the response-based KD, the model performance with feature-based and relationship-based KD needs to be further investigated. Third, compared to conventional model parameter-shared FL approaches, KD-based approaches solely rely on exchanging logits, resulting in a much smaller amount of information exchange. Therefore, despite they achieve model heterogeneity and require much less communication cost, the model performance may decrease. How to balance the model performance and the ability to accommodate model heterogeneity is a challenge. Finally, privacy concerns related to knowledge transmission need further exploration, especially given the ease with which some types of knowledge can be decoded and potentially accessed by unintended parties.

\subsection{PT-Based Approaches}

\subsubsection{Partial Training} PT refers to the strategy that allows clients to train specific segments of neural networks based on their computational capabilities. This methodology primarily divides into two branches, i.e., dynamic PT and post-PT. Dynamic PT focuses on training only a subset of the complete network, with the objective of generating diverse models in various FL scenarios while significantly reducing the computational demands. On the other hand, post-PT approaches aim at pruning the original full-size model on the server. The pruned model is subsequently fine-tuned on devices using their local data, a process initiated after acquiring the aggregated model. This approach strategically leverages local data to refine the model, thereby enhancing its applicability and effectiveness.

\subsubsection{Partial Training in FL} 
\begin{figure}[h]
    \setlength{\abovecaptionskip}{-0.cm}
    \setlength{\belowcaptionskip}{-0.cm}
    \centering
    \includegraphics[width=0.4\textwidth]{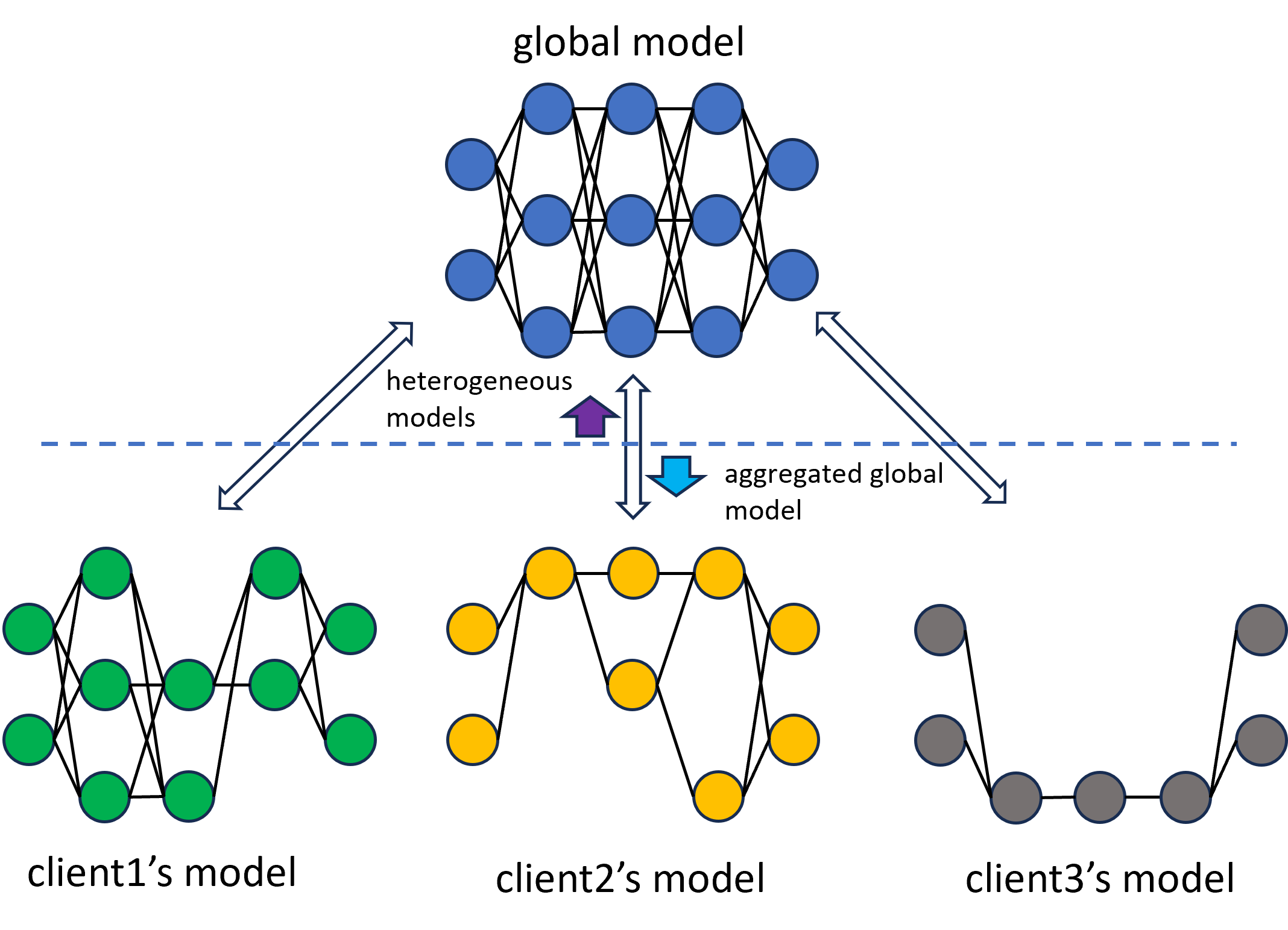}
    \caption{The illustration of PT in the model-heterogeneous FL.}
    \label{pt_based_FL}   
\end{figure}

Figure~\ref{pt_based_FL} illustrates the basic idea of PT in FL. Each client trains a part of the global model and only aggregates the corresponding parts during the aggregation step. HeteroFL~\cite{diao2020heterofl}, as a typical dynamic PT method, allows clients to train specific parts of the neural network based on their computational capacity and an adaptive communication compression scheme that adjusts model update compression according to clients' communication resources. In particular, the clients are organized into multiple groups based on their computation capabilities and communication resources. Each group independently trains its model locally using the PT technique, and the local models are then aggregated to form a global model. This method is suitable for HeteroFL since every communication round is independent. HeteroFL achieves significant improvements in terms of computation and communication efficiency on a benchmark dataset, making FL feasible for a wider range of devices and use cases. 

Compared to the static sub-model extraction strategy in HeteroFL, FedRolex~\cite{alam2022fedrolex} proposes a rolling sub-model extraction method. In this scheme, the sub-model is derived from the full-sized model using a progressively advancing rolling window with each communication round. This innovative strategy ensures that different sections of the global server model receive uniform training, thereby alleviating the client drift caused by model heterogeneity. In addition, while HeteroFL confines the global model's size to match that of its largest client model, FedRolex permits global models to surpass the complexity of any individual client model, leading to enhanced performance.

In contrast to HeteroFL and FedRolex dividing the clients into multiple groups, Hermes~\cite{Hermes} develops a two-stage technique that includes client selection and adaptive model compression, aiming to improve the efficiency of FL on diverse mobile clients. In the client selection stage, the authors employ a reinforcement learning algorithm to select the most suitable clients for participating in the training process based on their data quality, computational capabilities, and communication bandwidth. The selected clients then train their models locally using PT, and the updated model weights are sent to the central server for aggregation. In the adaptive model compression stage, the authors propose a novel approach that uses a combination of model pruning and quantization to compress the model size without sacrificing accuracy. This approach facilitates efficient transmission of model updates from the clients to the server, resulting in reduced communication costs. 

Heroes~\cite{10621351} adapts another perspective to achieve model-heterogeneous FL. It leverages the neural composition technique to construct size-adjustable models by composing low-rank tensors, where each model weight can be approximated by the product of two low-rank tensors called neural basis and coefficient. Models with heterogeneous complexities can be created by adjusting the size of the coefficient. During the training process, the clients download the basis and coefficient from the server and compose them into the local model for local training, then decompose them into basis and coefficient to send back to the server. Since the server may receive coefficients with different shapes, it performs a block-wise aggregation to update it. 

As a typical dynamic PT method, PriSM~\cite{niu2022federated} leverages the spatial structure of neutral networks, i.e., low-rank structure and networks kernel orthogonality to train sub-networks in the orthogonal kernel space. In particular, PriSM uses singular value decomposition on the original kernels to extract a set of principal orthogonal kernels. The significance of these kernels is determined by their singular values. Following this, it also introduces a unique sampling strategy that independently selects varying subsets of the principal kernels. This strategy is used to create client-specific sub-models, effectively reducing computational and communication demands. It is worth noting that kernels with larger singular values are given a higher likelihood of being selected during the sampling process. As a result, each sub-model represents a low-rank approximation of the comprehensive server model. 

Compared to dynamic PT approaches, post-PT approaches finetune the model by leveraging their local data. For example, LotteryFL~\cite{9708944} generates heterogeneous models, which is inspired by the Lottery Ticket hypothesis, presenting a method for identifying Lottery Ticket Networks (LTNs) – sparse subnetworks within larger models. In this context, LotteryFL aims to find each client's LTN during communication rounds in FL and only exchanges LTN parameters between clients and the server, thereby reducing communication overhead. In particular, LotteryFL employs client-specific pruning strategies to the updated weights, resulting in sparse subnetworks (winning tickets). In summary, the LotteryFL framework leverages the lottery ticket hypothesis to improve personalization and communication efficiency in FL for non-IID datasets. 

PruneFL~\cite{9762360} directly employs model pruning to create various local models to fit model-heterogeneous requirements. In particular, PruneFL proposes distributed pruning which involves initial pruning at a chosen client, followed by additional distributed pruning integrated within the standard FL process, enabling more efficient model compression across the network. Subsequently, PruneFL dynamically maintains a compact model with efficient transmission, computation, and reduced memory footprint, while preserving essential connections for convergence to an accuracy similar to the original model. In addition, PruneFL continuously updates the parameters to retain and adjust the model size, aiming to minimize the time taken to reach convergence. Experimental results demonstrate that this approach significantly reduces model size and communication costs while maintaining high accuracy on benchmark datasets. The PruneFL approach also shows improved efficiency in terms of training time and energy consumption, making it suitable for FL on resource-constrained edge devices. 

FlexFL~\cite{chen2024flexflheterogeneousfederatedlearning} is another PT-based approach that employs a flexible pruning strategy. Unlike most existing methods that predefine pruning ratios, FlexFL innovatively determines these ratios for each layer using the Activation Percentage of Zeros (APoZ), where a higher APoZ corresponds to a higher pruning ratio. Once the server generates heterogeneous models based on this mechanism, they are sent to the clients, allowing for further adaptive pruning based on the client's available resources. Additionally, FlexFL leverages KD to transfer knowledge from smaller models to larger ones, ensuring that even if most clients are resource-constrained, larger models are not undertrained.


In addition, Zhou et al.~\cite{zhou2024every} theoretically demonstrate the effectiveness of the PT methods. They propose a framework for model-heterogeneous FL methods with model extraction and present a convergence analysis to explore the theoretical guarantees of convergence on PT-based methods. The results prove that under certain conditions, these methods can converge to a stationary point of standard FL.

\subsubsection{Limitations} 
The PT methodology enables clients to tailor the training of specific sectors within neural networks, but it comes with several limitations. First, it is assumed that each local model is a subset of the global model, meaning they should share common elements. This requirement limits the types of local models, reducing flexibility in designing models for different local data. Second, most of the existing works employ either random or fixed strategies to obtain these partial neural networks. However, the optimal method to select the most crucial components for enhanced learning efficiency remains ambiguous. Third, while much of the existing research is grounded in empirical evidence, there is a notable lack of rigorous theoretical analysis.

\subsection{Alternative Approaches}
In addition to the KD and PT-based methodologies, several innovative techniques have been developed to enable model-heterogeneous FL. They either leverage knowledge from other areas of AI or propose entirely new methods from a mathematical perspective. 

HAFL-GHN~\cite{litany2022federated} handles heterogeneous client architectures by employing a Graph HyperNetwork (GHN). The framework does not require any public data and the client architectures remain private. GHNs are highly adaptable for accommodating any architecture, with their nodes embodying the layers of the client model and the edges of its computational workflow. Each local client initiates by training a copy of the GHN with its private data and then passes the resulting weights of the graph to the server. These weights are aggregated on the server side and distributed to local models for further optimization. HAFL-GHN outperforms the alternative frameworks based on KD or non-graph hypernetworks, especially when clients possess smaller local datasets. However, when the clients have homogeneous architectures, the performance results of the presented framework do not yet match that of other state-of-the-art methods. Another notable advantage of this method pertains to flexibility, i.e. allowing for changes in local clients or new clients can be added with minimal effort.

DISTREAL~\cite{rapp2022distreal} presents a Distributed Resource-Aware Learning pattern to address systems with various resources, allowing clients to adapt independently of the server. Such adaptation involves omitting dynamical filters of the convolutional layers to reduce the computational complexity of the training. This dropout mechanism is applied during runtime and uses a dropout rate, which is determined once before training and is defined by a Pareto-optimal vector. The rate is based on the convergence speed and the resource requirements and is determined by an automated design space exploration in the form of a genetic algorithm. Experimental results with different datasets have demonstrated that DISTREAL achieves significantly faster convergence speed in the FL system without sacrificing/compromising accuracy. This dynamic dropout approach maximizes resource utilization, allowing devices of varying specifications to participate seamlessly. 

~\cite{yao2021fedhm} propose FL for Heterogeneous Models (FEDHM) as a method to address the problem of heterogeneous models. Low-rank factorization is used to decompose the model parameters for different clients and divide them into shared and individual components. The shared components capture the common features across different models, while the individual components capture the unique features of each model. FEDHM applies FL to update the shared and individual components separately. This update approach for both shared and unique components reduces both communication costs and device training overhead, boosting performance across diverse settings. Empirical evidence highlights FEDHM's superior efficacy compared to the state-of-the-art methods. However, a notable drawback remains in the steep computational overhead linked to matrix composition.

\begin{table*}[htbp]
\small
\caption{A summary of the state-of-the-art model-heterogeneous FL methods.}
\label{summary}
\begin{tabular}{|>{\centering\arraybackslash}m{2cm}|>{\centering\arraybackslash}m{2.3cm}|>{\centering\arraybackslash}m{6cm}|>{\centering\arraybackslash}m{6cm}|}
\hline
\rowcolor[HTML]{C0C0C0} 
\textbf{Aggregation}        & \textbf{Method} & \textbf{Key Ideas} & \textbf{Limitations} \\ \hline
\end{tabular}
\begin{tabular}{|>{\centering\arraybackslash}m{2cm}|>{\raggedright\arraybackslash}m{2.3cm}|>{\raggedright\arraybackslash}m{6cm}|>{\raggedright\arraybackslash}m{6cm}|}
                   &  FedMD~\cite{li2019fedmd} &   Conduct KD on public labeled dataset to achieve aggregation & Heavily relies on public datasets, limited performance                   \\ \cline{2-4} 
                           &  FedDF~\cite{lin2020ensemble}   &  Use unlabeled dataset to conduct KD on the server side & Still rely on public dataset, even if it is unlabeled    \\ \cline{2-4} 
                           &  FedGen~\cite{zhu2021data} &  Use data generator to achieve data-free KD & Training data generator introduces extra costs \\ \cline{2-4} 
                           & FedFTG~\cite{Zhang_2022_CVPR} &  Propose a hard sample mining strategy to prompt the generator to produce samples that align more with the real distributions &  Introduce extra cost for the data generation process                 \\ \cline{2-4} 
                           & FedProto~\cite{tan2022fedproto} &  Employ class prototype as a medium for knowledge exchange  &  Only valid in computer vision tasks \\ \cline{2-4}
                           & FedHKD~\cite{chen2023best} &  Introduce hyper knowledge to improve performance in heterogeneous data, do not rely on public dataset  & Similar to FedProto, rely on class prototype, valid only in computer vision tasks  \\ \cline{2-4} 
                           &  Fed-CMA~\cite{regatti2022conditional} &  Align class conditional distributions of clients in feature space  &  The impact of model heterogeneity level is unclear \\ \cline{2-4}
                           &  FedTSA~\cite{fan2024fedtsaclusterbasedtwostageaggregation} &  Leverage a diffusion model to generate public dataset, offering high-quality synthetic data and good flexibility  &  Only applicable to image classification tasks \\ \cline{2-4}
                           &  DFRD~\cite{wang2024dfrd} &  Train a conditional generator consisting of three losses: fidelity loss, transferability and diversity  &  Too many losses lead to higher computation overhead \\ \cline{2-4}
                           &  FedGH~\cite{wang2024dfrd} &  Split the model into a shared prediction header and a personalized feature extractor, only the headers are aggregated and updated  &  Sensitive to data heterogeneity  \\ \cline{2-4}
                           \hline
\multirow{-26}{*}{KD-based} & HeteroFL~\cite{diao2020heterofl}  &  Each client trains part of the whole model and only the overlapping parts are aggregated  &  Non-overlapping parts cannot be updated, leading to potential bias  \\ \cline{2-4} 
                           &  FedRolex~\cite{alam2022fedrolex}  &  Introduce a rolling mechanism to ensure all parts have the same chance to be updated  &   Lack convergence analysis, cannot guarantee convergence  \\ \cline{2-4} 
                           & Hermes~\cite{Hermes} & Each device trains a subnetwork by applying structured pruning and only the overlapped parameters are aggregated   & Like HeteroFL, the non-overlapping parameters never get updated   \\ \cline{2-4} 
                           &  PriSM~\cite{niu2022federated} &  Use singular value decomposition to extract principal orthogonal kernels to produce sub-models that can be aggregated   & The method is limited to CNN structure \\ \cline{2-4} 
                           & LotteryFL~\cite{9708944} &  Each client learns a LTN and only these networks will be exchanged between clients and server  &  The LTNs are unstructured sparse, which does not significantly reduce computational cost \\ \cline{2-4} 
                           & Hereos~\cite{10621351} &  Effectively balance communication efficiency and model accuracy in heterogeneous edge networks  &  Introduce complexity in model design and implementation, hard to deploy in real-world applications \\ \cline{2-4} 
                           & FlexFL~\cite{chen2024flexflheterogeneousfederatedlearning} & Propose a APoZ-guided flexible pruning strategy  &  Pruning may lead to potential information loss \\ \cline{2-4}
\multirow{-18}{*}{PT-based} & PruneFL~\cite{9762360} & Employ adaptive and distributed parameter pruning to create heterogeneous models &  Lack comparison with other model-heterogeneous FL baselines  \\ \hline
                           & DISTREAL~\cite{rapp2022distreal} & Propose a dropout mechanism that dynamically adjusts the model complexity & Limited to CNNs \\ \cline{2-4} 
                           & FEDHM~\cite{yao2021fedhm} & Adopt low-rank factorization to decompose the model into shared and individual components & High computational cost for decomposition \\ \cline{2-4}
\multirow{-6}{*}{Others}   &  HAFL-GHN~\cite{litany2022federated}  &  Aggregate heterogeneous models by a Graph HyperNetwork &  The architecture needs to be predefined, limiting scalability \\ \hline
\end{tabular}
\end{table*}

\section{Open problems and Future Directions}
\label{open_problem}

Model-heterogeneous FL is an emerging yet practical distributed learning paradigm. Despite the considerable research on this topic, we identify several open problems and future directions that demand further research. 

\subsection{Aggregation Strategy}

In model-heterogeneous FL, traditional element-wise weight averaging for aggregation becomes impractical due to differences in model architectures, affecting the efficacy of many homogeneous FL techniques. To address this, existing methods either use latent representation to standardize the models, like employing KD for logit processing, or employ pruning to ensure models have overlapping sections for aggregation. However, both approaches have inherent challenges. KD-based methods assume that client logits are aligned, either from a shared dataset or synthetic data. When this alignment is lacking, logit aggregation can lose its relevance, as the logits may represent entirely different features. In pruning-based methods, focusing only on overlapping components may neglect essential, non-overlapping model features, resulting in suboptimal model performance.

We envision promising future research directions toward model aggregation strategy. First, while KD strategies often rely on logits, this is constrained by data alignment. A more promising direction might be to investigate relation-based knowledge~\cite{huang2022knowledge,dong2021few}, which is calculated by the inner products between features across layers. Such knowledge contains the implicit learning logic of a model, which is irreverent to the input data, eliminating the reliance on public datasets. Regarding pruning-based strategy, an interesting direction is to identify the principal components, in other words, identify the layers and weights that most impact model performance, akin to principal component analysis (PCA)~\cite{wold1987principal,}. With a clear understanding of the principal components, efficient pruning techniques can be employed to modify model architectures, while preserving the most important features of a model.

\subsection{Privacy and Security}
Despite FL enabling data to remain local, sharing model parameters can still potentially expose private information. An adversary might reverse-engineer the raw data from model information, as both weights and gradients are derived from the input data through specific matrix operations~\cite{10574838}. In model-heterogeneous FL, the transmitted data includes either output logits or a subset of model parameters, leaving it unclear whether transferring this information mitigates the risks of reverse engineering. Model poisoning is another significant security concern in FL systems~\cite{10423783,10386784}. A malicious actor could manipulate the updated model before sending it to the server for aggregation, degrading the performance of the global model. In model-heterogeneous FL, where clients use different types of models, direct modifications to the model have a limited impact on the global model, as the update strategy does not rely solely on model parameter aggregation. However, quantifying this impact is crucial for designing more robust and efficient model-heterogeneous FL methods.

\subsection{Resource-constrained Devices}
Model-heterogeneous FL offers a promising solution for involving resource-constrained devices in FL. However, low-tier devices such as embedded controllers, cameras, and electric meters often lack the capability to train even the simplest neural networks. Additionally, the computational cost of local training can significantly increase power consumption, potentially disrupting the core functions of these devices. Therefore, effectively deploying model-heterogeneous FL on low-tier devices remains an open challenge. One potential direction is offloading parts of the training tasks from resource-constrained devices to trusted edge nodes using split learning~\cite{10138067,10234718}. Another approach is to explore more advanced FL training strategies, such as allowing clients with limited computational capacity to train simpler models (e.g., logistic regression), while others train more complex models like neural networks.

\subsection{Model-heterogeneous FL on Big Data}
With the increasing amount and diversity of data across various domains, applying FL to Big Data presents significant challenges. Traditional FL methods, which require clients to collaboratively train a unified global model, struggle to address the heterogeneity in both data distribution and system resources. Model-heterogeneous FL allows data producers to customize local models tailored to their specific data characteristics and system constraints, greatly improving training effectiveness. This approach also lowers the barriers to sharing Big Data in a privacy-preserving manner. However, Big Data scenarios are often highly dynamic, with data continuously being generated and overall distributions constantly shifting. Additionally, the diversity of data types necessitates the integration of multimodal learning. As a result, developing real-time FL algorithms and effectively handling multimodal scenarios~\cite{10.1145/3580305.3599825} are promising directions for future research.

\subsection{Model-heterogeneous FL in the Generative AI Era}
With the development of foundation models and multimodal technologies, Generative AI has demonstrated phenomenal performance across various domains~\cite{epstein2023art}. However, the privacy issues of training data for downstream tasks and the massive scale of foundational models (e.g., GPT~\cite{brown2020language}, Llama~\cite{touvron2023llama}) hinder further development of Generative AI. In this scenario, model-heterogeneous FL emerges as a promising solution, offering a viable pathway to mitigate concerns around data privacy and computational requirements. By adopting model-heterogeneous FL, small-scale parties lacking substantial computational resources can also contribute to training smaller-scale models in a privacy-preserving manner, lowering barriers to entry for foundation model research. Moreover, model-heterogeneous FL can also combine with parameter-efficient fine-tuning (PEFT) methods to further improve usability and flexibility. For example, arranging low-rank adaptation (LoRA)~\cite{hu2021lora,cho2024heterogeneouslorafederatedfinetuning} modules with different ranks based on the resources of clients. Such a strategy not only optimizes resource utilization but also paves the way for personalized and efficient model fine-tuning, significantly expanding the potential applications of Generative AI in a manner that is both resource-conscious and privacy-aware.

\section{Conclusion}
\label{conclusion}
This paper presents a comprehensive review of model-heterogeneous FL methods. We start with discussing the challenges posed by conventional FL, setting the background for understanding the motivations behind model-heterogeneous FL. Subsequently, we dive into the state-of-the-art model-heterogeneous FL strategies, classifying them into two primary directions, i.e., KD-based and PT-based approaches. We also highlight notable methods that do not fit neatly into these categories. Drawing from comparative insights into these methods, we illustrate the pros and cons of each method and spotlight unresolved challenges and potential directions for future research. This paper serves as a structured guide to model-heterogeneous FL and underscores its importance in real-world scenarios.

\bibliographystyle{IEEEtran}
\bibliography{main}

\end{document}